# Experimental control of $T_c$ in AlB$_2$-type compounds using an applied voltage

Jose A. Alarco, Mahboobeh Shahbazi and Ian D.R. Mackinnon
(Queensland University of Technology, Brisbane Australia)

We utilise the van de Pauw technique combined with Density Functional Theory to show that an external voltage applied to superconducting AlB$_2$-type compounds, such as MgB$_2$, and Mg(B$_{1.9}$C$_{0.1}$), effects substantive changes in transition temperature due to modification of electronic band structure(s). An applied voltage results in a symmetric split of degenerate σ bands in MgB$_2$, similar to that calculated for atom displacement along the B–B bond aligned with E$_{2g}$ phonon mode directions. For AlB$_2$, similar splitting of σ bands occurs albeit with higher voltage requirement compared to MgB$_2$. Experimental data show that $T_c$ values for MgB$_2$ and Mg(B$_{1.9}$C$_{0.1}$), reduce consistently with increased increments of applied voltage. Zero resistance is limited by an upper applied voltage depending on the compound.

correspondence: ian.mackinnon@qut.edu.au

**Introduction**   AlB$_2$-type superconductors show properties consistent with the Bardeen-Cooper-Schrieffer (BCS) model[1] and behave in a manner in between that of semi-conductors and metals. The boron-boron (B–B) bonds of MgB$_2$ are clearly implicated in superconductivity and euphemistically described as covalent bonds driven metallic[2]. Such description reflects observations that extrinsic factors are key to fundamental mechanisms of superconductivity[3] and hence, may be manipulated to change macroscopic properties[4]. Well-known experimental practices to control or influence the superconducting transition temperature, $T_c$, include doping or substitution of suitable element(s)[5] and application of mechanical stress (*e.g. via* hydrostatic pressure)[6].

Models on mechanical stress aligned to specific direction(s) in AlB$_2$-type structures are proposed as a means to effect superconductivity[7] in phases such as ScB$_2$. Other computational studies[8,9] explore electron doping and tensile strain as a potential means to control $T_c$ and superconducting topologies. These studies emphasise orientation specific use of strain/stress to drive changes in hole- or electron-doped bands in AlB$_2$-type electronic structures[10]. A recent study that benchmarks methods to assess external factors (*i.e.* stress, strain) influencing the $T_c$ of MgB$_2$ provides intriguing suggestions on the future use of strain engineering to moderate superconductivity[11]. In this work, we show by calculation and confirm by experiment that changes in $T_c$ value of superconducting AlB$_2$-type compounds can be achieved using an applied voltage.

Experiments and first principles calculations by Pogrebnyakov *et al.*,[12] showed that epitaxial strain in MgB$_2$ films on appropriate substrates increase $T_c$ by up to 1.5 K above that for an unstrained sample. The tensile strain in MgB$_2$ is due to reduction in cell parameters with increased film thickness resulting in a softening of the E$_{2g}$ mode[12]. The E$_{2g}$ phonons dominate the electron-phonon coupling in MgB$_2$ [2] and Density Functional Theory (DFT) calculations show splitting of degenerate σ bands along the Γ–A reciprocal direction (*i.e.*, along the *c*- or $c^*$-axis) with B–B bond stretching in the *a-b* plane[2,13]. This change in electronic structure is mirrored by changes to the Kohn anomaly in the phonon dispersion (PD) for MgB$_2$ [14].

We outline below an additional method to adjust and, perhaps to control, the $T_c$ of AlB$_2$-type structures and we link application of voltage, or an electric field, to the breakdown of superconductivity. We examine the effects of an external voltage applied to AlB$_2$-type superconductors *via in situ* experiment and *via* (DFT) calculation.

**Methods**   General methods for sample preparation and characterisation as well as DFT calculations are outlined in earlier publications[15-17] and Supplemental Material for methods and sample nomenclature. Pellets of MgB$_2$ and Mg(B$_{1.9}$C$_{0.1}$) produced in-house, pressed and sintered to form consolidated pieces are used in this study. The operational practice for physical property measurements is equivalent to the van der Pauw four-point probe configuration used for Hall effect and resistivity[18]. In this configuration, a reverse polarity (or reverse current) option reduces thermal drift and exploits combined signals to deliver high quality output[18]. The accuracy of the van der Pauw technique has been evaluated for a range of configurations including for thick samples of regular or irregular morphology[19,20]. For this work, the thickness of sample(s) is less than the lateral dimensions and contacts are at the edges of the sample to ensure accuracy and reproducibility[19]. For each sample, an increment of 0.1V is applied after each temperature decrease/increase sweep above and below the superconducting transition temperature range.

Inspection of individual channels during measurements using the van der Pauw configuration show that the level of current within samples does not exceed 12 mA at temperatures above $T_c$ for these compounds. We estimate the average cross section of a sample to be ~ 4–5 mm$^2$, given a sample thickness of 1 mm and contact distance ~3 mm. With these physical dimensions, the normal state current is ~0.3 A/cm$^2$. This applied current is a value substantially lower than the predicted critical current > $10^4$ A/cm$^2$ based on magnetic measurements and the Bean model for polycrystalline $MgB_2$ [21]. To further minimize the potential for transport critical current, conducting bridge(s) or weak link(s) were avoided during measurements of resistivity. The observed voltage effects on resistive transitions using the van der Pauw configuration appear to be solely electronic band structure effects.

The CASTEP option of the Materials Studio 2021 software[22] used to calculate the effect of an applied potential on the electronic band structure is identifiable under the CASTEP "Electronic Options" Tab. An electric field can be specified in magnitude of eV/Å/e and along specific X, Y, Z directions. This calculation requires the symmetry of the structure to be reduced to P1. The structure is then geometry optimised from the input geometry to determine the predicted effects of an applied external voltage on the geometry of the crystal and on the EBS. The same starting cell parameters, functional(s) and computational conditions are used for systematic comparisons of DFT outcomes.

Voltage Effects on $T_c$ – Experimental: The influence of an externally applied voltage using the van der Pauw configuration utilises the same experimental strategy and measurement configuration to evaluate the effect of a change in voltage on $T_c$ for each sample. The $T_c$ values for all samples reduce consistently for increased increments of applied voltage as shown in Figure 1. The signal to noise of the resistance response is poor at low applied voltage (*i.e.* << 0.2 V) and improves with higher applied voltage up to a limiting value which differs for each composition.

The $T_c$ values for $MgB_2$ reduce consistently for increased increments of applied voltage. Resistance measurement(s) become unstable or inconsistent after an upper applied voltage value of 3.4 V as shown in Figure 1a. A similar result with lower upper voltage of 2.2V is obtained for sample 2-$MgB_2$ (see Supplemental for data). The upper voltage value is dependent on the volume density of material and limits to current carrying capacity (*e.g.* local heating), or other factors such as substantial change in structure or electronic properties under these experimental conditions.

For $MgB_2$ at high applied voltage (*i.e.*, >2.4 V), resistance values for each temperature sweep (decrease/increase) are identifiable as shown in Figure 1a. A detailed plot at applied voltages between 2.4 V and 3.2 V is available in Supplemental Materials to highlight these different resistance(s) for each sweep. The difference in value between onset $T_c$ and zero $T_c$, $\Delta T_c$, for $MgB_2$, is 0.23 K at 0.1 V. This value for $\Delta T_c$ is consistent with earlier detailed studies on $MgB_2$ [23][24]. The range of measurable onset $T_c$ values for 0 < V < 3.4 V is equivalent to ~1.2 meV. A plot of onset and zero $T_c$ values with applied voltage for $MgB_2$ in Figure 1a is shown in Supplemental Materials. See also Figure S3a in Supplemental Materials showing resistance data with temperature for sample 2-$MgB_2$ prepared under different conditions.

Carbon-doped $MgB_2$ (*i.e.* $Mg(B_{1.9}C_{0.1})$) at zero magnetic field also shows consistent reduction in onset and zero $T_c$ for increased increments of applied voltage. The signal to noise of the resistance response substantially improves with higher applied voltage and reaches a limit at which the response is inconsistent or at higher resistance than at zero applied voltage below $T_c$. Figure 1b shows resistance data for $Mg(B_{1.9}C_{0.1})$ using the van der Pauw configuration for applied voltages 0.2 V < V < 2.2 V (sample 2- $Mg(B_{1.9}C_{0.1})$ in Table S1). For carbon-doped $MgB_2$, the difference between onset and zero $T_c$ is greater than that for $MgB_2$ and ranges from 1.7(1) K to 2.9(1) K at higher applied voltages (*i.e.* > 1.0 V). Figure S3 in Supplemental Materials shows the same experiment for a second sample of Mg(B$_{1.9}$C$_{0.1}$) showing a lower value of the upper voltage (1.2 V) after which resistance increases substantially. This sample also shows inconsistent resistance behaviour with temperature sweep(s). For example, the data for 0.3 V and 0.4 V in Figure S3 show substantially different resistance profiles with temperature sweep. At a similar value for resistance along the vertical slope (*e.g.,* 5x10$^{-6}$ ohm), the difference in increase/decrease sweep for these two applied voltages is between 0.03 K and 0.08 K, respectively.

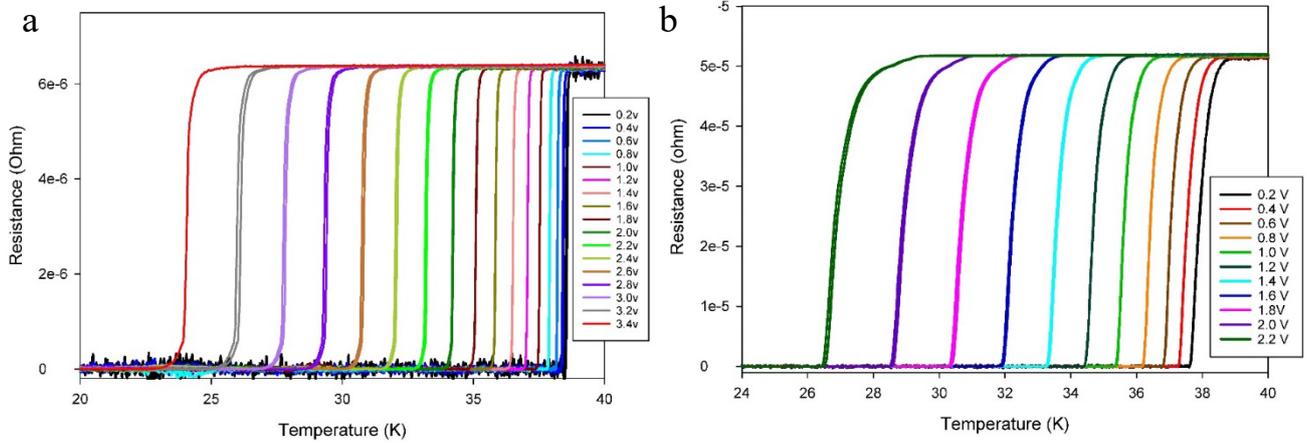

Figure 1: Resistance measurements for (a) sample 1-MgB$_2$ and (b) sample 2-Mg(B$_{1.9}$C$_{0.1}$) with applied external voltage using a van der Pauw configuration[20] for measurements. Note the successive reduction of onset and zero T$_c$ with incremental increases in applied voltage. The temperature range of the superconducting transition (*i.e.*, onset to zero T$_c$) increases with increased applied voltage.

Calculated Voltage Effects – EBS: In DFT models for AlB$_2$-type structures, we emphasize specific bands within 1 eV to 2 eV of the Fermi level because these bands reflect variations to hybrid bonding orbitals and transitions to/from valence and conduction bands[25]. We show the influence of an applied external field in a specific direction of the unit cell (*i.e.* along the *b*-axis, designated V$_y$) using DFT calculations of the EBS for MgB$_2$ in Figure 2a. Notice that the σ-bands, implicated in superconductivity of MgB$_2$ [2,26], display substantial splitting or degeneracy breaking, similar to effects previously calculated for displaced atomic positions along directions corresponding to E$_{2g}$ modes[2,13].

While Figure 2a shows three different types of band distributions, each condition of voltage and direction has been independently calculated and overlain on the same plot. With increase in an applied external field, V$_y$, the σ bands remain split, or non-degenerate, for calculated applied voltages up to V$_y$ = 0.14 eV/Å/e at which point the lower band is parallel with the Fermi level at Γ (*i.e.* m$_L$, the light effective mass, band value is -1meV at Γ). Note that while the σ bands are split at Γ under these applied voltage conditions, the trajectories of the bands parallel the V$_y$=0.0 eV/Å/e bands above and below the Fermi level (depending on value of V$_y$) on either side of Γ.

Further increases in the applied voltage results in expansion of the energy gap between the split σ bands with the lower band crossing below the Fermi level at Γ. For example, at V$_y$ = 0.2 eV/Å/e, the σ bands remain split with the lower band 136 meV below the Fermi level at Γ. The π band at M shows an increase in energy relative to the Fermi level with an applied external field in V$_y$. In comparison to the σ band at Γ, the π band increases to 355 meV above the Fermi level when V$_y$=0.14 eV/Å/e (compared with an equilibrium value of 173 meV for V$_y$=0). The π band is at higher energy (403 meV) when V$_y$=0.2 eV/Å/e (*i.e.* when the light effective mass, m$_L$, band is below the Fermi level).

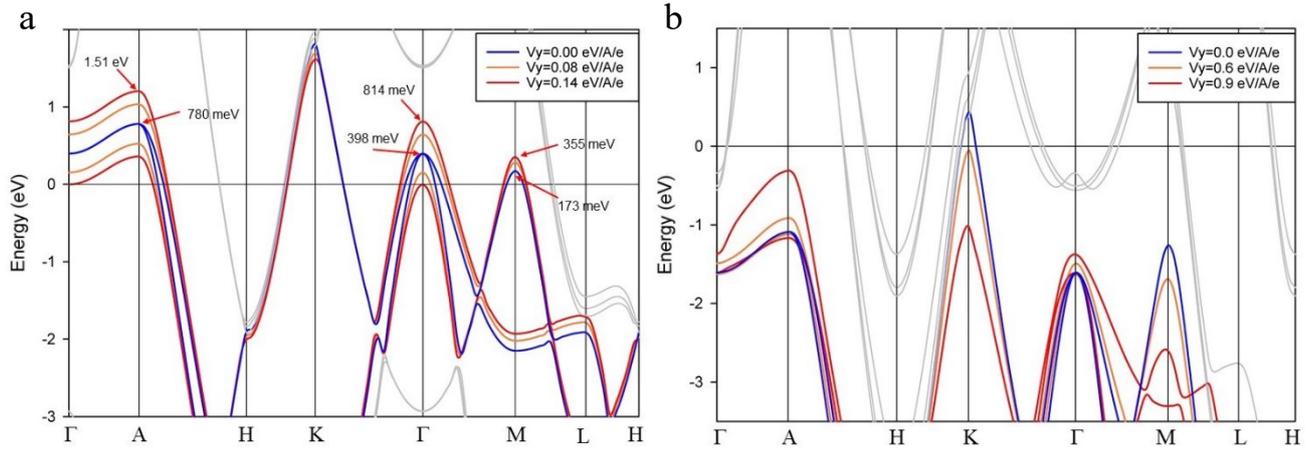

Figure 2: Calculated EBS for (a) $MgB_2$ and (b) $AlB_2$ with different values of applied external voltage ($V_y$; as eV/Å/e) showing transition to non-degenerate σ bands with increased voltage. In both cases, with increased effect along the $b*$ direction, the σ bands along the axial direction(s) (*i.e.* along Γ–A) are non-degenerate. For $MgB_2$, at $V_y = 0.14$ eV/Å/e the non-degenerate lower σ band (*i.e.* the light effective mass, $m_L$, band) is at the Fermi level.

Figure 2b shows the effect of applied voltage on the calculated EBS for $AlB_2$. Figure 2b shows that the effect on $AlB_2$ for $V_y = 0.6$ eV/Å/e and $V_y = 0.9$ eV/Å/e results in non-degenerate σ bands. The threshold voltage at which the σ bands are clearly non-degenerate is $V_y > 0.2$ eV/Å/e. At higher applied voltage(s), the difference between the upper and lower bands increases particularly along Γ–A, but notably, these bands are substantially less sensitive to applied voltage than those for $MgB_2$.

A larger comparative shift in energy occurs for the bands at K and M in the reciprocal lattice towards lower energy levels. The trajectories of bands at K and at M shift to lower energy with applied voltage(s) $V_y < 0.5$ eV/Å/e (data not shown). At $V_y > 0.6$ eV/Å/e, splitting of the σ band is readily apparent. For $V_y$, the energies of bands at K and M undergo substantial reduction from the equilibrium position (*i.e.* at $V_y = 0.0$ eV/Å/e) and substantial distortion to σ band dispersion along Γ–A at higher applied voltage.

**Discussion** The $MgB_2$ suite of superconducting materials (*i.e.*, $(Mg_{1-x}M_x)(B_{1-y}C_y)_2$ where M=Al, Sc, Li, Ti) are well-documented examples of phonon-mediated superconductivity, readily accessible for computational and experimental evaluation[17] and potential manipulation of band structure(s)[2,14]. These analytical tools have been used to suggest ways to increase $T_c$ for $AlB_2$-type and similar isovalent and/or isostructural compounds[27] using the substitutional, mechanical or stress/strain related methods noted above. To date, we have not detected specific evidence nor predictions in the literature for modification of superconductor behaviour, particularly $T_c$, using an applied external voltage or electric field.

In an insightful summary of nanoelectronics developments at the time, Yang *et al.*[28] highlight advantages of the electric-field control of lattice, charge, orbital, and spin degrees of freedom in novel materials. While their survey focused on the multiferroic $BiFeO_3$, the fundamental energy-scale(s) and time-scale(s) that influence these degrees of freedom (see Figure 1 of[28]) also apply to superconductors. The nanoelectronics industry has demonstrated that electric field control of physical properties is possible and can be fine-tuned to suit specific purpose(s)[28]. Consideration of this mechanism for superconductors may be prospective for a targeted $T_c$ in $AlB_2$-type materials. This type of manipulation is implied for other superconductors as recently shown for the $Cd_3As_2$ Dirac semimetal nanowire Josephson junction using gate tuning to shift from electron to hole conduction[29].

Resistance Measurements: The superconductors in this study – $MgB_2$ and $Mg(B_{1.9}C_{0.1})$ – show similar effects on the value(s) of $T_c$ with application of voltage, or an applied electric field. Figure 3a shows the change in $T_c$ with applied voltage for these samples as well as values for $T_c$ determined by the zero field cooled (ZFC) method. These values are different to the $T_c$ values obtained by the van der Pauw technique by 0.5 K and 1.5 K due to the different configurations required for such determinations. Nevertheless, the trends showing a reduction in $T_c$ with an applied voltage are clearly shown in Figure 3a.

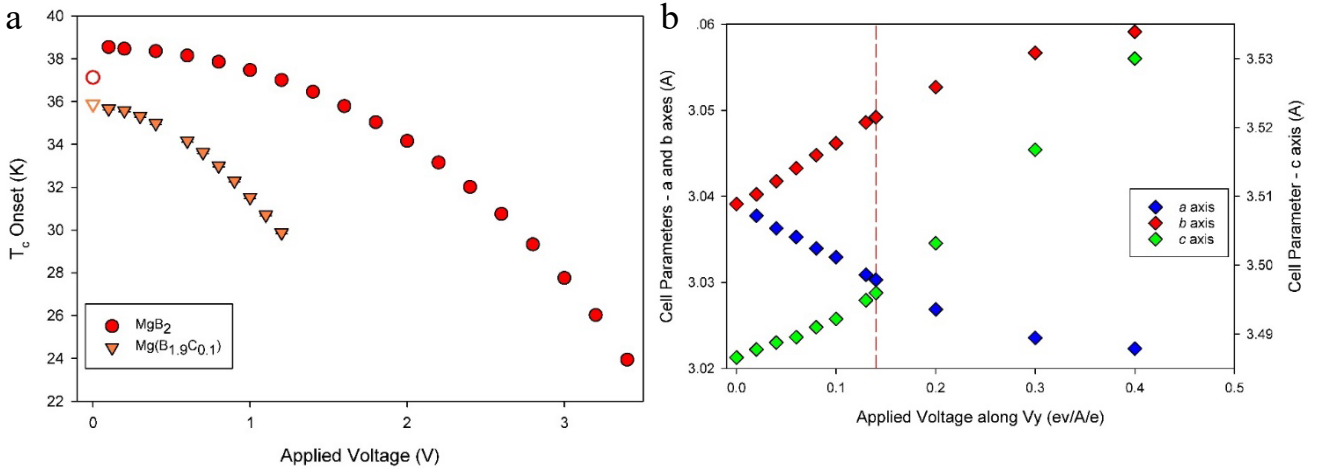

Figure 3: (a) Plot of experimentally determined $T_c$ (onset) for $MgB_2$ and $Mg(B_{1.9}C_{0.1})$ superconductors with applied external voltage using the van der Pauw technique[20]. For each data point, the standard deviation in $T_c$ is within the symbol size. Open symbols are values for ZFC $T_c$ determined at V=0; and (b) Calculated cell parameters for $MgB_2$ with applied voltage along the Y-axis (co-incident with the *b*-axis in the initial P6/mmm unit cell). Values for *a*- and *b*-axes are on the left-hand ordinate and for the *c*-axis on the right. Dotted vertical line in (b) indicates the voltage for which the calculated light effective mass band is parallel with the Fermi level at Γ.

We attribute these changes in $T_c$ to the influence of an applied electric field on bonding and, thus on cell dimensions and electron-phonon behaviour. In particular, we propose that an electric field applied parallel to the *a–b* plane substantially influences B–B bond lengths. Similarly, a voltage applied perpendicular to, or at a consequential angle to the *a–b* plane (*e.g.* along [110] or [111]), also results in variations to cell dimensions to a greater or lesser extent. The extent of change to cell dimensions – or to bonding characteristics – is dependent on the amount of applied voltage and the orientation of the respective electric fields produced. Consequently, the cell dimensions of $MgB_2$ and other superconductors subject to similar applied voltage(s) will vary and may be fine-tuned to achieve greater or lesser extent of electron- or hole-doping in electronic band(s).

Subtle changes to crystal lattice(s) and crystal symmetry[14,15,25] are fundamental mechanism(s) that lead to superconductivity and to adjustment of physical properties such as $T_c$. Electronic properties of $AlB_2$-type structures are affected by changes in bonding and/or bond lengths particularly as electron- or hole-doping is effected either by substitution into the metal[30] or the boron plane (*e.g.* carbon;[10]) or metal non-stoichiometry[31]. As shown with DFT calculations for unsubstituted $MgB_2$, a change in B–B overall bond length of < 0.063 Å (*i.e.* ~0.03 Å either side of the equilibrium position) along the $E_{2g}$ mode direction(s) induces significant changes to the PD and Fermi surfaces[13].

Within single crystal or polycrystalline samples of $MgB_2$, changes to lattice dimensions are readily apparent with change(s) in temperature[32], particularly with neutron or X-ray diffraction techniques[5,24,32,33]. In general, these changes in cell dimensions from room temperature to below 40 K range from 0.08 % to 0.18 % for the *a* and *c* axes, respectively, when referred to the P6/mmm space group. These temperature induced changes result in a shift in the average B–B bond length of 0.0015 Å[32] to 0.0025 Å[24], respectively, depending on the precision and method of determination.

The shift in the B–B bond length with temperature is substantially less than that demonstrated for dynamic variation along the $E_{2g}$ direction for the hole-doped $MgB_2$ superconductor[2,13]. In DFT analyses, the deformation potential for $MgB_2$ at ~40 K is equivalent to a bonding shift of ~0.06 Å along the *a–b* plane[13], ~20 times that induced by reducing the temperature from room temperature to a superconducting state < 40 K. Hence, we infer that the B–B bond, or electron density, is capable of substantial shifts within the *a–b* plane coincident with phonon behaviour in the $E_{2g}$ directions[26]. This inference is consistent with earlier commentary that subtle shifts in crystal structure and/or symmetry occur as $MgB_2$ (and other $AlB_2$-type superconductors) are brought to $T_c$ [3,14,25].

While in the superconducting state, electron-phonon behaviour is moderated by the form of Fermi surfaces represented by hole-doped σ bands that cross the Fermi level to/from the valence bands. This shift in the width and form of Fermi surfaces along the *c*- or *c\**-axis direction is well described for $MgB_2$ with pressure[34] and for $(Mg_{1-x}Al_x)B_2$ with Al content[35]. In these works, the inner Fermi surface parallel to the Γ–A direction "pinches out" (or is reduced to zero) when superconductivity – effected by increased hydrostatic pressure or change in Al content – is lost[34,35]. We propose that similar variations in cell dimensions and consequent distortions of Fermi surfaces are induced by applied voltage(s) on $AlB_2$-type superconductors with resulting effect on $T_c$ as shown in Figure 1.

Applied Voltage Calculations: Similar effects are calculated for the EBS of $AlB_2$-type structures when subjected to an applied voltage. Different effect(s) are due, in part, to (i) the orientation(s) of the applied voltage in relation to the reciprocal/real lattice directions of the modelled compound, (ii) the specific direction of application (*e.g.* whether along the *y*-axis; *z*-axis or all three axes), (iii) the magnitude of applied voltage in relation to the physical size/shape of measured material and (iv) whether the compound is a single crystal or a compacted powder. Computational outcomes are premised on a single crystal format albeit we are not able to confirm these data with experiment on a singe crystal of $MgB_2$ at this time. Our comparisons with experimental data refer to outcomes for bulk compacted powders.

Figure 3b shows the variation of cell parameters for $MgB_2$ when subjected to an applied voltage along the Y-direction that corresponds to the [010] direction of the asymmetric unit (a P1 cell derived from P6/mmm symmetry). Application of voltage in a direction aligned with the *b*-axis, results in an increase of both the *b* and *c* cell parameters although the *a* parameter shows reduced dimension. With increased applied voltage, the calculated cell volume increases by ~0.5% (compared to the equilibrium condition) for $V_y$=0.14 eV/Å/e when the $m_L$ band is at the Fermi level.

The calculated EBS for $MgB_2$ with an applied voltage along the *b*-axis shows that in general, the cell volume increases with increased voltage and that the structure shifts from the typical P6/mmm symmetry attributed to the equilibrium condition to a P1 symmetry. While this is clearly a theoretical construct, DFT principles would suggest that the substantive changes in cell dimensions largely mirror response(s) to an applied electric field on hybrid bonding orbitals in the *a–b* plane. The vertical dashed line in Figure 3b delineates the approximate value of $V_y$ for which the $m_L$ band is parallel with the Fermi level at Γ. At applied voltage(s) higher than this delineation, we expect that the compound does not superconduct and/or is unstable. We suggest that this condition is shown experimentally when resistance measurements increase substantially above the background value obtained at lower voltages and when the signal is unstable.

Figure 4 shows the Fermi surface of $MgB_2$ calculated with the application of 0.1 eV/Å/e in the Y-direction. The inner Fermi surface (darker blue colour in Figure 4a) necks in the middle, around the Γ point when viewed approximately along [100]. This change in Fermi surface form, or necking, resembles the form calculated with application of pressure [34] or with B–B bond deformation along the $E_{2g}$ directions in $MgB_2$ [13]. However, calculations of the Fermi surfaces when viewed along the [001] direction, show a substantial difference to $MgB_2$ without an applied voltage.

For example, at the boundary zone near Z in reciprocal space, Fermi surfaces have lost the typical concentric projection as shown in Figure 4b (see Figure 7 of[13] and Figure 5 of[34] for Fermi surface calculations of $MgB_2$ without an applied voltage). A loss of form for these two critical Fermi surfaces (*i.e.* non-concentric parallel planes) reduces the extent of reciprocal space within which resonance between adjacent Fermi surfaces may be coupled to $E_{2g}$ phonons[13,34]. Under these conditions, we would anticipate a gradual reduction in $T_c$ with systematic application of an external electric field as shown in Figure 1. This calculated effect on Fermi surfaces differs from the known influence of electric fields on Fermi surfaces attributed to electronic transport properties in metals. In the case of metals, rigid displacements of Fermi surfaces are expected for fields perpendicular to the surface and no displacement is expected for fields parallel to the Fermi surfaces[36].

Calculated cell dimensions for $AlB_2$ with applied voltage consistent with DFT methods for $MgB_2$ are provided in Supplemental Materials. For $AlB_2$, the change in cell dimensions does not mirror that shown in Figure 3b in that the *a*-axis does not substantially reduce in dimension compared to the $V_y$=0.0 condition. With applied voltage, both the *b*-axis and *c*-axis dimensions increase substantially and by similar proportions from the equilibrium condition for $AlB_2$. This shift in cell dimensions also presupposes a symmetry change to P1 but with asymmetric split bands in the EBS as shown in Figure 2b. In contrast, calculations suggest that $AlB_2$ requires a higher applied voltage to produce structural changes or distortions of comparable magnitude to those induced in $MgB_2$. In contrast to implications by Cheng *et al.*,[8] these calculated outcomes suggest that enhanced electronic properties for $AlB_2$, for example, by use of an applied voltage are difficult to implement.

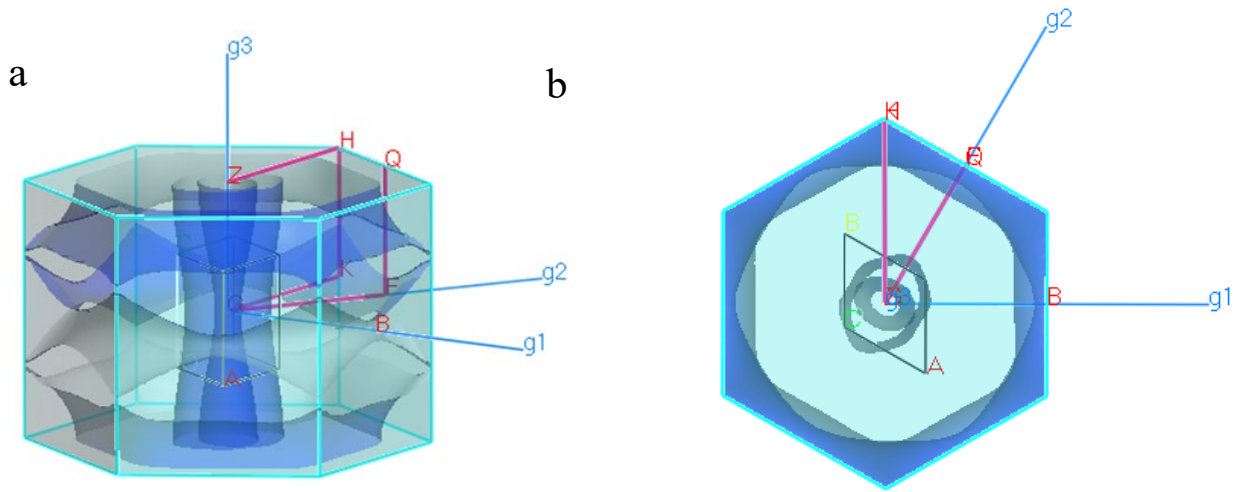

Figure 4: Calculated Fermi surfaces for $MgB_2$ after application of 0.1 eV/Å/e in the Y-direction (a) viewed perpendicular to the *c\**-axis and (b) along the *c\**-axis. Note that in (a) the inner Fermi surface necks inward towards Γ and that in (b) the Fermi surfaces at Z are no longer concentric with application of an electric field.

Degenerate Electronic Bands: We have shown for a range of stable, single metal $AlB_2$-type structures, that only $MgB_2$ shows a combination of σ bands and π bands above the Fermi level in an EBS as well as a definitive Kohn anomaly along the Γ–M and Γ–K reciprocal directions in a converged phonon dispersion (PD) model[14,16,17]. For $MgB_2$, the σ bands along the Γ–A direction (*i.e.* along the *c*- or *c\**-axis) are degenerate and approximately parallel the Fermi level (a "flat band"). For $AlB_2$, $ScB_2$, $YB_2$ and $TiB_2$, electronic bands along the Γ–A direction and at Γ, are degenerate; but only for $ScB_2$ and $YB_2$ is some portion of these bands in the Γ–A direction above the Fermi level[14]. For other single metal diborides, only $ScB_2$ shows a similar, but much diminished, combination of features in the EBS and PD and, consequently, a lower experimentally determined value for $T_c$ at ~1.5 K[37].

Experiments and calculations on $AlB_2$-type structures highlight that degenerate bands, particularly along the Γ–A direction and at Γ (along the Γ–K and Γ–M directions), are inimical to the structure. However, these degenerate bands are only hole-doped for those $AlB_2$-type structures that superconduct. These bands are key to understanding electronic properties defined by valence and conduction band states. For example, with increased metal or carbon substitution into $MgB_2$, either the degenerate σ bands in the EBS are electron filled or substantial shifts occur in the frequency and order of the $E_{2g}$ phonons in the PD. Either or both effects result in a lowered $T_c$ compared with pure $MgB_2$ or in complete loss of superconductivity. Degenerate EBS bands may be beneficial for superconductivity in AlB$_2$-type compounds, but they may not be a necessary requirement for superconductivity.

Two key effects identified by DFT modelling – bond deformation energy[2,13] and an applied voltage – induce a split of degenerate bands for AlB$_2$-type structures. In both cases, modelling shows similar variations in the form and direction of split bands albeit the influence on valence bands, for example with AlB$_2$, is of lower relative magnitude. The σ bands of superconducting compounds remain split with increased atom displacement or applied voltage to a condition where the m$_L$ band is at or below the Fermi level. At this point, superconductivity is non-existent and the topology of coaxial, nearly parallel "tubular" Fermi surfaces in the *c*- or *c\**-axis direction is not observed. We infer that the split of degenerate σ bands in the EBS of superconducting AlB$_2$-type compounds does not destroy superconductivity, *per se*, but that a degree of latitude enables electron-phonon interaction while coaxial Fermi surfaces are present along the Γ–A direction.

In summary, we have shown *via* DFT calculations that an external voltage applied to MgB$_2$ and to Mg(B$_{1.9}$C$_{0.1}$) will influence their electronic properties as shown by splitting of degenerate σ bands in the EBS. These split σ bands are indicators of changes to electronic properties that we suggest are manifest in superconducting properties. Experimental confirmation for this is demonstrated by a systematic reduction in T$_c$ with application of an applied voltage using the van de Pauw four-point probe technique. Calculations show that π bands are also influenced by an applied voltage. For superconducting forms, π bands increase in energy and remain above the Fermi level as conduction bands with increase in applied voltage. For AlB$_2$, an applied voltage results in an asymmetric split of degenerate bands and a strong shift of valence bands at the M reciprocal node to lower energies. The full impact of these observations is yet to be determined, but we suggest this approach offers potential for precise manipulation of superconductor electronic properties.

**Acknowledgements**
This research did not receive any specific grant from funding agencies in the public, commercial or not-for-profit sectors. Access to, and ongoing assistance with, QUT's HPC facilities particularly through the e-Research Office and access to experimental equipment via the Centre for Clean Energy Technologies and Practices as well as the Central Analytical Research Facility are gratefully acknowledged.

**References**

[1] J. Bardeen, L.N. Cooper, and J.R. Schrieffer, Phys. Rev. **108**, 1175 (1957).
[2] J.M. An and W.E. Pickett, Phys. Rev. Lett. **86** (19), 4366 (2001).
[3] A. Bianconi, J Supercond Nov Magn **33**, 2269 (2020).
[4] K.W. Lee and W.E. Pickett, Phys. Rev. Lett. **93** (23), 237003 (2004); K.-Y. Yeh, Y.-R. Chen, T.-S. Lo, P.M. Wu, M.-J. Wang, K.-S. Chang-Liao et al., Frontiers in Physics **8**, 1 (2020).
[5] J. Karpinski, N.D. Zhigadlo, G. Schuck, S.M. Kazakov, B. Batlogg, K. Rogacki et al., Phys. Rev. B **71** (174506), 1 (2005).
[6] S. Deemyad, T. Tomita, J. Hamlin, B. Beckett, J. Schilling, D. Hinks et al., Physica C: Superconductivity **385** (1), 105 (2003); A. Bianconi and T. Jarlborg, Europhysics Letters **112** (37001), 1 (2015).
[7] H. Zhai, F. Munoz, and A.N. Alexandrova, Jour. Mater. Chem. C **7**, 10700 (2019).
[8] C. Cheng, M.-Y. Duan, Z. Wang, and X.-L. Zhou, Philosophical Magazine **100** (17), 2275 (2020).
[9] C. Cheng, M.-Y. Duan, W.-X. Xu, Z. Wang, and X.-L. Zhou, Philosophical Magazine **101** (24), 2599 (2021).
[10] D. Kasinathan, K.-W. Lee, and W.E. Pickett, Physica C **424**, 116 (2005).
[11] E. Johansson, F. Tasnadi, A. Ektarawong, J. Rosen, and B. Alling, Jour. App. Physics **131** (063902), 1 (2022).
[12] A.V. Pogrebnyakov, J.M. Redwing, S. Raghavan, V. Vaithyanathan, D.G. Schlom, S.Y. Xu et al., Phys. Rev. Lett. **93** (147006), 1 (2004).
[13] J.A. Alarco, P.C. Talbot, and I.D.R. Mackinnon, Mod. Numer. Sim. Mater. Sci. **8**, 21 (2018).
[14] J.A. Alarco and I.D.R. Mackinnon, in *Phonons in Low Dimensional Structures*, edited by Vasilios N. Stavrou (InTech Open, London UK, 2018), pp. 75.

[15] J.A. Alarco, A. Chou, P.C. Talbot, and I.D.R. Mackinnon, Phys. Chem. Chem. Phys. **16**, 24443 (2014).

[16] J.A. Alarco, P.C. Talbot, and I.D.R. Mackinnon, Phys. Chem. Chem. Phys. **17** (38), 25090 (2015).

[17] I.D.R. Mackinnon, Almutairi A., and J.A. Alarco, in *Real Perspectives of Fourier Transforms and Current Developments in Superconductivity*, edited by Juan M.V. Arcos (IntechOpen Ltd., London UK, 2021), pp. 1.

[18] M. Reveil, V.C. Sorg, E.R. Cheng, T. Ezzyat, P. Clancy, and M.O. Thompson, Review of Scientific Instruments **88** (094704), 1 (2017).

[19] C. Kasl and M.J.R. Hoch, Review of Scientific Instruments **76** (033907), 1 (2005).

[20] S.H.N. Lim, D.R. McKenzie, and M.M.M. Bilek, Review of Scientific Instruments **80**, 1 (2009).

[21] I.D.R. Mackinnon, A. Winnett, J.A. Alarco, and P.C. Talbot, Materials **7** (5), 3901 (2014).

[22] S.J. Clark, M.D. Segall, C.J. Pickard, P.J. Hasnip, M.I.J. Probert, K. Refson et al., Z. Kristallogr. **220**, 567 (2005).

[23] J.D. Jorgensen, D.G. Hinks, and S. Short, Physical Review B **63** (22) (2001).

[24] S. Lei, H. Zhang, L. Chen, and Y. Feng, Journal of Physics: Condensed Matter **16** (36), 6541 (2004).

[25] J.A. Alarco, B. Gupta, M. Shahbazi, D. Appadoo, and I.D.R. Mackinnon, Phys. Chem. Chem. Phys. **23** (41), 23922 (2021).

[26] J.M. An, S.Y. Savrasov, H. Rosner, and W.E. Pickett, Phys. Rev. B **66** (220502), 1 (2002).

[27] H. Rosner, A. Kitaigorodsky, and W.E. Pickett, Phys. Rev. Lett. **88** (127001) (2002); W.E. Pickett, J.M. An, H. Rosner, and S.Y. Savrasov, Physica C **387**, 117 (2003).

[28] J.-C. Yang, Q. He, P. Yu, and Y.-H. Chu, Annu. Rev. Mater. Res. **45**, 249 (2015).

[29] W.R. Li, H.S. Ye, J. Sheng, J.J. Jiang, B.Y. Shen, Z.Y. Li et al., IEEE Trans. Appl. Superconductivity **30** (4), 1 (2020).

[30] A. Bianconi, S. Agrestini, D. Di Castro, G. Campi, G. Zangari, N.L. Saini et al., Phys. Rev. B **65** (17), 174515 (2002).

[31] I. Loa, K. Kunc, K. Syassen, and P. Bouvier, Phys. Rev. B **66** (13), 134101 (2002).

[32] J.D. Jorgensen, D.G. Hinks, and S. Short, Phys. Rev. B **63** (224522), 1 (2001).

[33] J. Karpinski, N.D. Zhigadlo, S. Katrych, R. Puzniak, K. Rogacki, and R. Gonnelli, Physica C **456**, 3 (2007); J. Karpinski, N.D. Zhigadlo, S. Katrych, K. Rogacki, B. Batlogg, M. Tortello et al., Physical Review B **77** (214507), 1 (2008).

[34] J.A. Alarco, P.C. Talbot, and I.D.R. Mackinnon, Physica C: Supercond. and Applications **536**, 11 (2017).

[35] O. de la Penha, A. Aguayo, and R. de Coss, Phys. Rev. B **66**, 01251 (2002).

[36] J.M. Ziman, *Principles of the Theory of Solids*, Second ed. (Cambridge University Press 1972).

[37] S.M. Sichkar and V.N. Antonov, Low Temperature Physics **39** (7), 595 (2013).

# Supplemental Data

Experimental control of $T_c$ in $AlB_2$-type compounds using an applied voltage

Jose A. Alarco, Mahboobeh Shahbazi and Ian D.R. Mackinnon
Queensland University of Technology, Brisbane Australia

**Methods:** Pellets of $MgB_2$ and $Mg(B_{1.9}C_{0.1})$ produced in-house, pressed and sintered to form consolidated pieces are used in this study. Powder samples are synthesised by established solid-state methods using starting materials and heating protocols described in earlier publications [1]. Rectangular pieces (~ 6mm in lateral dimensions; thickness ~ 1 mm) are attached to sample holders by direct contact with silver paint for physical property measurements [2]. Measurements are made with a Cryogenics Ltd Mini System, equipped with closed cycle cooling to 1.5 K and an external magnetic field up to 5 Tesla.

The van der Pauw configuration and methodology provide internally consistent resistivity and temperature data from synthesised samples. $T_c$ values for onset and for zero resistance are determined from an average of six temperature values that encompass order of magnitude change(s) in resistance. Estimates of average $T_c$ and standard deviation are based on resistance values obtained at ~20 second intervals. Zero field cooled data for samples are also provided for comparison. Experiments with applied voltages are duplicated on two $MgB_2$ and two $Mg(B_{1.9}C_{0.1})$ samples synthesised under different conditions using the method noted above. Details of synthesis conditions and data on these samples are provided in Table S1 below.

Polycrystalline samples are characterised by powder x-ray diffraction (XRD) using Co Kα1 radiation in Bragg Brentano geometry with 0.02° 2θ steps and a counting time of 10 s per step using a Bruker D8 Advance X-ray diffractometer. Diffraction patterns are refined and indexed using the software program Topas; quantitative estimates of phase abundance in each product are determined by Rietveld refinements using Topas. In general, phase abundances determined by this technique are within <5% relative error [3]. Phase analysis data and refined cell dimensions for the superconducting compounds are also provided in Table S1.

Software used is the CASTEP module of the Materials Studio 2021 package [4]. We have shown that computational accuracy and precision for DFT models not only depend on the type of functional utilised [5] but also on the appropriate use of key parameters such as the k-grid interval and the energy cut-off value [6]. This degree of computational resolution is important for modelling meV-scale phenomena such as superconductivity in $AlB_2$-type structures [6,7]. In this work, we show data for DFT calculations using a k-grid value of 0.01 Å$^{-1}$ (*i.e.*, a grid size ~ 38x38x29; or 20,938 k-points), an energy cut-off of 990 eV and the LDA functional.

**Other Data:**

**Table S1**
Parameters for metal diboride syntheses and experimental data (standard deviations in parentheses)

| Sample ID | $MgB_2$ (%) | MgO (%) | $T_{max}$ (°C) | $T_{hold}$ (min) | a (Å) | c (Å) | $T_c$ (K) via ZFC | $T_c$ (K) (onset) | $T_c$ (K) (zero) |
|---|---|---|---|---|---|---|---|---|---|
| 1-$MgB_2$ | 95.45 | 4.55 | 800 | 30 | 3.0867(2) | 3.5241(3) | 37.13 | 38.55(6) | 38.32(5) |
| 2-$MgB_2$ | 96.76 | 3.24 | 850 | 120 | 3.08595(2) | 3.52531(3) | 37.30 | 38.74(10) | 38.31(10) |
| 1-$Mg(B_{1.9}C_{0.1})$ | 95.57 | 4.43 | 650 | 30 | 3.08550(2) | 3.52300(2) | 35.89 | 35.67(6) | 35.37(6) |
| 2-$Mg(B_{1.9}C_{0.1})$ | 96.61 | 3.39 | 850 | 60 | 3.08188(1) | 3.52661(2) | 36.30 | 38.55(3) | 37.52(4) |

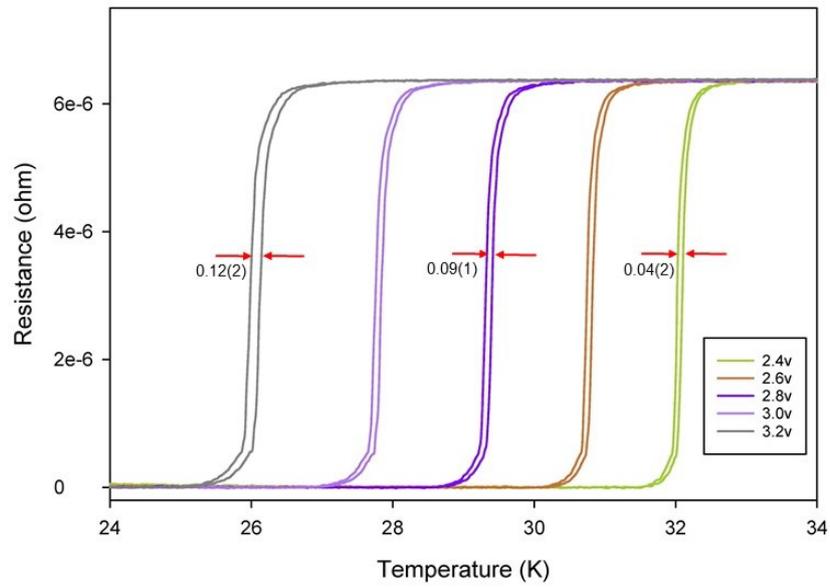

Figure S1: Detailed view of data shown in Fig. 1a for applied voltages between 2.4 V and 3.2 V. The difference in temperature for the increasing and decreasing temperature sweeps increases with applied voltage.

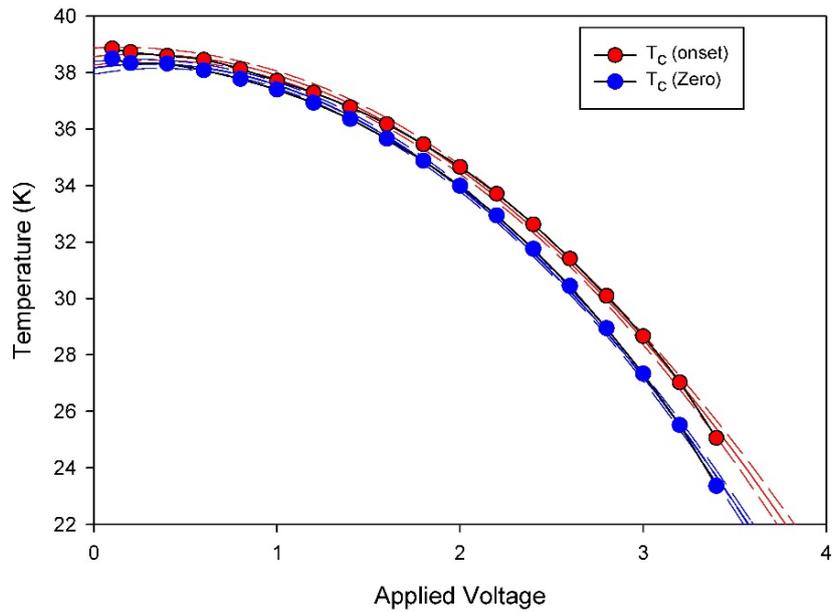

Figure S2: Plot of onset and zero $T_c$ determined using the van de Pauw technique for polycrystalline sample of $MgB_2$. Dotted lines delineate 95% confidence levels for a polynomial fit with $R^2=0.99$.

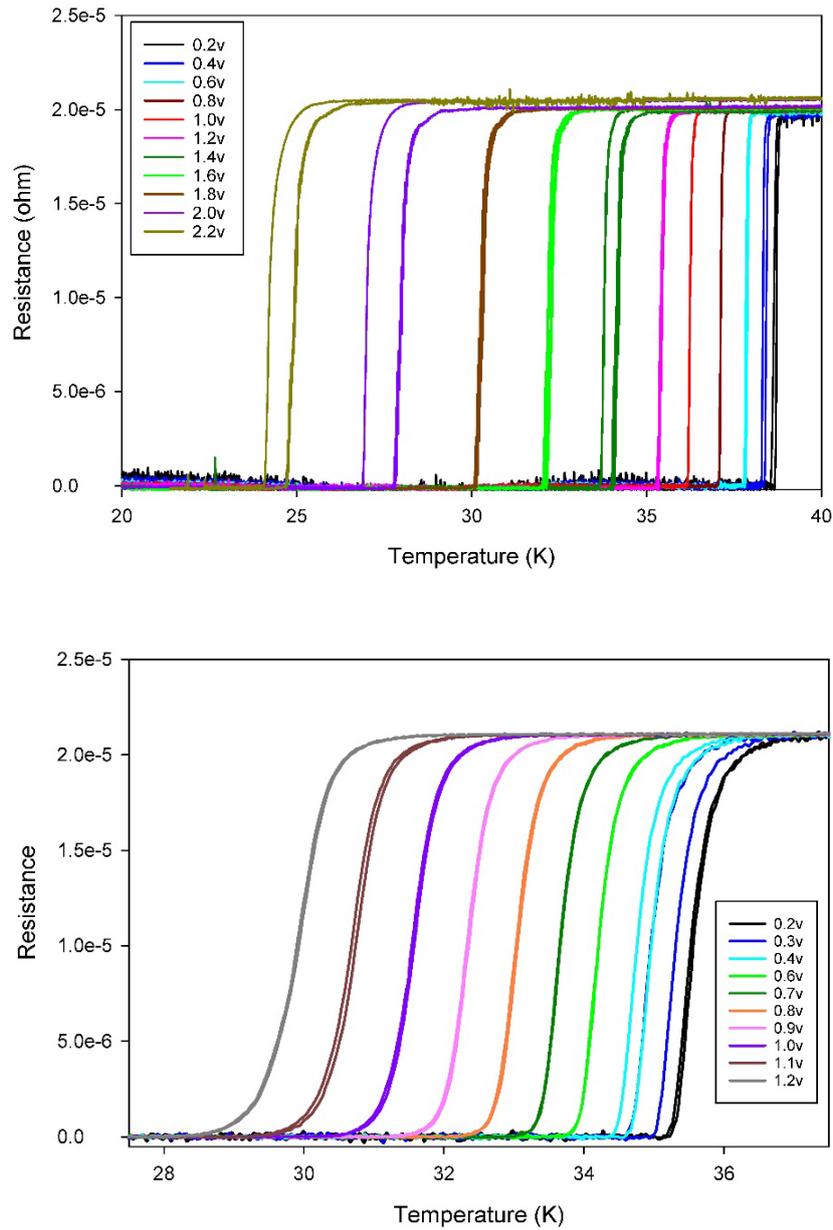

Figure S3: Resistance measurements for samples (a) 2-MgB$_2$ and (b) 1-Mg(B$_{1.9}$C$_{0.1}$) (Table S1) with applied external voltage using a van der Pauw configuration [8] for measurements. In both cases, the applied voltage range for superconducting behaviour is less than that for respective measurements shown in Figure 1 due, in part, to different synthesis conditions and degree of effective sintering of the consolidated pellets. Note that for 2-MgB$_2$, each temperature sweep (increase/decrease) is evident for applied voltages of 1.4, 2.0 and 2.2 volts, respectively. The "noise" in Figure S3a at near zero resistance and the higher resistance values before T$_c$ (onset) for higher applied voltage (e.g. $2.0 \leq V \leq 2.2$) suggests conduction pathways across and between grain boundaries are more problematic under these conditions.

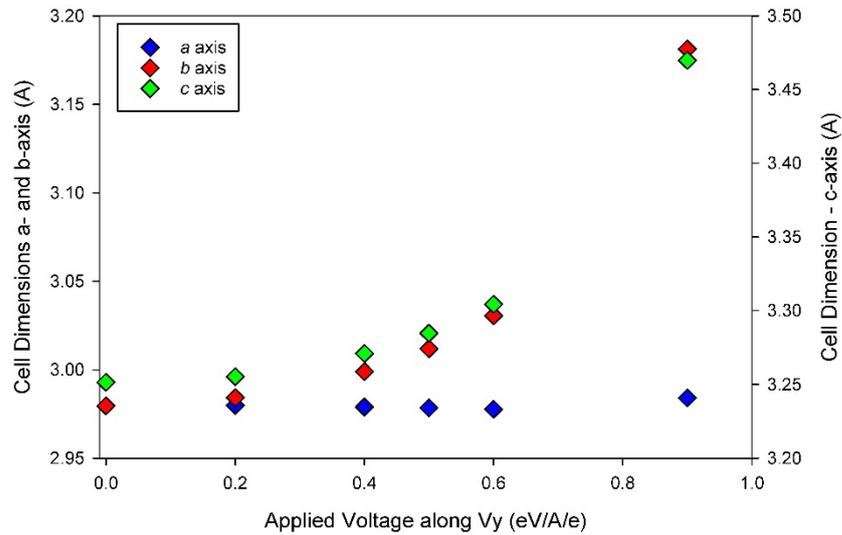

Figure S4: Calculated cell parameters for AlB$_2$ with applied voltage along the Y-axis (co-incident with the *b*-axis in the initial P6/mmm unit cell). Values for *a*- and *b*-axes are on the left-hand ordinate and for the *c*-axis on the right.

**References:**

[1] M. Shahbazi, A. Dehghan-Manshadi, M. Hossain et al., IEEE Transactions on Applied Superconductivity **31** (6200305), 1 (2021).

[2] C. Kasl and M.J.R. Hoch, Review of Scientific Instruments **76** (3) (2005).

[3] L.B. McCusker, R.B. Von Dreele, D.E. Cox et al., J. Appl. Crystallogr. **32** (1), 36 (1999).

[4] S.J. Clark, M.D. Segall, C.J. Pickard et al., Z. Kristallogr. **220**, 567 (2005); K. Refson, P.R. Tulip, and S.J. Clark, Phys. Rev. B **73** (155114), 1 (2006).

[5] J.A. Alarco, P.C. Talbot, and I.D.R. Mackinnon, Mod. Numer. Sim. Mater. Sci. **4**, 53 (2014).

[6] I.D.R. Mackinnon, Almutairi A., and J.A. Alarco, in *Real Perspectives of Fourier Transforms and Current Developments in Superconductivity*, edited by Juan M.V. Arcos (IntechOpen Ltd., London UK, 2021), pp. 1.

[7] D. Kasinathan, K. Koepernik, J. Kunes et al., Physica C-Superconductivity and Its Applications **460**, 133 (2007).

[8] S.H.N. Lim, D.R. McKenzie, and M.M.M. Bilek, Review of Scientific Instruments **80** (7) (2009).